\documentstyle[12pt]{article}
\input epsf
\title{Nucleation of vacuum phase transitions by topological defects}

\author{William A. Hiscock\\Department of Physics\\Montana State
University\\Bozeman, Montana  59717, U.S.A.}

\begin{document}
\maketitle
\vspace{5cm}
\begin{abstract}

The Euclidean action is calculated in the thin-wall approximation for a
first-order vacuum phase transition in which the bubble appears symmetrically
around either a global monopole or a gauge cosmic string. The
bubble is assumed to be much larger than the core size of the monopole or
string. In both cases the
value of the Euclidean action is shown to be reduced below the $O(4)$
symmetric action value, indicating that the topological defects act as
effective nucleation sites for vacuum decay.
\end{abstract}

\thispagestyle{empty}
\newpage

Over the last two decades, vacuum phase transitions have evolved from being
an oddity in electroweak theory to being an expected part of the early
universe, likely responsible for inflation, and possibly the source of
exotic topological structures such as domain wall, strings, and monopoles.
In addition, there is also the disquieting possibility that we live in
a metastable false vacuum state today, the decay of which would be the
ultimate natural catastrophe.

The Euclidean action approach to calculating vacuum decay rates was first
developed in the thin-wall approximation by Coleman \cite{Co}. While subsequent
investigation has shown that the thin-wall approximation is only truly valid
for a very small set of potentials \cite{SaHi}, it is always valid in the
ultimate limit wherein the true and false vacuum states are nearly degenerate
in
energy. There are two ways in which this near degeneracy may be achieved in the
real universe. First, as the universe expands and cools, every potential will
pass through a state of degeneracy, due to the thermal terms in the
potential; for a broad variety of potentials, the phase
transition from false to true vacuum will take place while the potential is
nearly degenerate. In this case, the thin-wall approximation may still be
sufficiently valid to yield an adequate approximation to the true Euclidean
action. Secondly, if the false and true vacuum are truly nearly degenerate
at zero temperature (which results in an extremely long decay time for the
false vacuum--conceivably even as long as $10^{10}$ years) then the thin-wall
approximation is obviously valid.

In this letter the Euclidean action for a
thin-wall bubble nucleating symmetrically around a global monopole or a
gauge cosmic string is calculated. The spacetime metric is taken to be
that appropriate for the exterior region of a global monopole or gauge string
and is fixed. The gravitational effects of the vacuum energy density are
ignored; they will not substantially influence the value of the action unless
the symmetry breaking scale is comparable to the
Planck energy \cite{CoDeL}. The energy scale of the vacuum phase transition is
assumed to be
substantially lower than that which is responsible for the topological defect;
this assures that the core region of the monopole or string is only a
negligible
fraction of the nucleating bubble's interior volume. The symmetry breaking
which
created the global monopole or cosmic
string is assumed to be independent of the phase transition under
consideration;
Preskill and Vilenkin have considered the opposite case in which the vacuum
phase transition affects the symmetry groups whose quotient defines the
topological defects \cite{PrVi}. The presence of the global monopole or gauge
cosmic string reduces the symmetry of the nucleating bubble solution from the
usual $O(4)$ to $O(3)$ or $O(1)$, respectively.

The Euclidean action for bubbles nucleated around monopoles and strings is
found to be less in both cases than the empty space $O(4)$ action. This implies
that these objects will act as nucleation sites for vacuum phase transitions,
hastening the decay of the false vacuum. The decrease in the action is
proportional to the deficit angles associated with the global monopole or
cosmic string; since these are anticipated to be small compared to unity
(perhaps of order $10^{-6}$), the decrease in the action is small. Nucleation
of
the vacuum phase transition by a topological defect could still be important
in cases where the metastable false vacuum has an extremely long lifetime;
this is demonstrated at the end of this Letter.

Might the gravitational field of other types of topological defects nucleate
vacuum phase transitions? The external field of a gauge monopole is simply
the Schwarzschild metric with the monopole mass as the mass parameter. For
symmetry breaking scales sufficiently close to the Planck mass, gauge monopoles
can be significantly gravitationally bound; the quantity $M\over r$ has the
value $\eta^2$ at the "surface" of the monopole, defined as the Compton
wavelength of the monopole, $1\over \eta$, where $\eta$ is the energy scale of
the
symmetry breaking which creates the monopoles. If the gauge monopole is
sufficiently gravitationally bound, then the curvature of spacetime can
substantially decrease the Euclidean action \cite{MeHi}. However, the
reduction in action is a function of $M/R_0$, where $R_0$ is the bubble
nucleation radius; the monopole will then only have a significant effect
if the nucleating bubble were about the same size as the monopole,
which would require that the energy scale of the phase transition
be of the same order as of that which
created the monopole. This is unlikely. The gravitational
field of a global string \cite{CoKa} might reduce the action of a vacuum
bubble;
however, the metric is sufficiently complicated in algebraic form so that
analysis is difficult. The analysis of possible nucleation by domain walls
or textures is complicated by the non-static nature of those defects, which
makes a Euclidean approach problematic.

The metric describing the exterior of a gauge cosmic string was first derived
in the context of the linearized Einstein equations by Vilenkin \cite{Vi}; his
solution was later shown to be exact \cite{Go,Hi1,Li} and unique \cite{Hi1}
in the full nonlinear theory of general relativity. The exterior spacetime
is exceedingly simple, being merely flat space with a wedge excised, i.e. a
cone. The metric has the form
\begin{equation}
	ds^2 = -dt^2 + d\rho^2 + dz^2 + {C_s}^2\rho^2d\phi^2 ,
\end{equation}
where ${C_s}^2=1-8\pi\mu$, and $\mu$ is the mass per unit length of the string,
$\mu \approx \eta^2$, where $\eta$ is again the energy scale of the symmetry
breaking which creates the topological defect.
The deficit angle of the cone is thus $8\pi\mu$. This spacetime is truly flat,
with vanishing curvature everywhere off the axis of symmetry. For purposes of
treating an (almost) spherically symmetric bubble, it is convenient to recast
the cosmic string metric into spherical-style coordinates,
\begin{equation}
	r = (\rho^2+z^2)^{1 \over 2},
\end{equation}
\begin{equation}
	\theta = \tan^{-1}({\rho \over z}),
\end{equation}
which yields the metric form
\begin{equation}
	ds^2 = -dt^2 + dr^2 + r^2d\theta^2 + {C_s}^2 r^2 \sin^2(\theta) d\phi^2 ,
\end{equation}
The approximate metric describing the spacetime geometry outside the core of
a global monopole has been derived by Barriola and Vilenkin \cite{BaVi}. It may
be written in the form
\begin{equation}
	ds^2 = -{C_M}^2 d{\bar t}^2 + {C_M}^{-2} d{\bar r}^2 + {\bar r}^2
	(d\theta^2 + \sin^2(\theta) d\phi^2) ,
\end{equation}
where
\begin{equation}
	{C_M}^2 = 1-8\pi\eta^2.
\end{equation}
The metric of Eq.(2) represents a set of scalar fields whose mass function
$M({\bar r})$ is growing linearly with radius. In terms of a mass defined by
Keplerian orbits at arbitrarily large radii, the global monopole actually has
infinite mass. The metric takes on a more appealing form if new time and radial
coordinates are defined by
\begin{equation}
	t = {C_M}{\bar t} ,
\end{equation}
\begin{equation}
	r = {C_M}^{-1}{\bar r} ,
\end{equation}
which leads to:
\begin{equation}
	ds^2 = -dt^2 + dr^2 + {C_M}^2 r^2 (d\theta^2 + \sin^2(\theta) d\phi^2).
\end{equation}
In this form, the global monopole metric may be seen to describe Minkowski
space
with a deficit solid angle of $8\pi\eta^2$. While similar to the gauge cosmic
string metric, it should be emphasized that the global monopole spacetime is
not flat; in fact, the divergent mass can cause large tidal effects far from
the monopole core \cite{Hi2}.

The equation of motion for the bubble wall may be easily obtained from
conservation of energy. The total energy of the bubble consists of two pieces,
the wall energy and the interior energy, and must be zero. The bubble wall
rest and kinetic energy takes the form
\begin{equation}
	4\pi C^2 \sigma R^2 (1+\dot R^2)^{1 \over 2} ,
\end{equation}
where $\sigma$ is the bubble wall density, $R$ is the bubble wall radius,
and $\dot R$ is the proper time derivative, ${d \over {d \tau}}$
of the bubble wall radius.
The interior energy is the difference between the bubble interior in the
false vacuum state and the true vacuum state,
\begin{equation}
	-{{4\pi C^2} \over 3} \epsilon R^3 ,
\end{equation}
where $\epsilon$ is the false vacuum energy density minus the true vacuum
energy
density. In both Eqs.(10) and (11), the factor $C^2$ represents either
${C_s}^2$
or ${C_M}^2$, for the cases of the gauge string or global monopole,
respectively.
This factor appears in the energy of the bubble wall and interior when the
integrals over the $\theta$ and $\phi$ coordinates in Eqs.(4) and (9). The
total energy, the sum of the expressions in Eqs.(10) and (11), may then
be solved for ${\dot R}^2$,
\begin{equation}
	{\dot R}^2 = \left ( {\epsilon R} \over {3 \sigma} \right) ^2 - 1,
\end{equation}
which is precisely the same algebraic form as in the usual empty space $O(4)$
calculation \cite{Co}; the $C^2$ factor from the string or monopole has
canceled out of the equation of motion for the bubble.

The false vacuum decay rate $\Gamma$ has the form \cite{Co}
\begin{equation}
	\Gamma = A \exp (-B) ,
\end{equation}
where $A$ is a complicated factor involving a functional determinant, and
$B$ is the difference between the Euclidean action of the true-vacuum bubble
solution and the action of the false vacuum background. The factor $A$ has
units of inverse four-volume for $O(4)$ symmetric decay; $\Gamma$ then gives
the
number of bubbles nucleated per unit four volume. In the case of nucleation
along a cosmic string, $\Gamma$ will give the number of bubbles nucleated per
unit length of string per unit time; $A$ then has dimensions of inverse
length squared. For nucleation around a global monopole (or any other $O(3)$
symmetric situation), $\Gamma$ give the number of bubbles nucleated per unit
time; $A$ has dimensions of inverse length.

The calculation of the Euclidean action is straightforward. Rotating to a
Euclidean time coordinate, $\tilde t = it$, reverses the sign of ${\dot R}^2$,
\begin{equation}
	{\dot R}^2 = 1 -\left ( {\epsilon R} \over {3 \sigma} \right) ^2,
\end{equation}
where now an overdot represents differentiation with respect to Euclidean
proper
time, $\tilde{\tau}$. The bubble thus forms at $\tilde t = \tilde{\tau} = 0$
at an initial
radius $R_0 = {{3 \sigma} \over \epsilon}$, and then collapses down to $R = 0$;
Eq.(14) may be directly integrated to find the dependence of $R$ on
$\tilde{\tau}$:
\begin{equation}
	R = R_0 \cos \left ( {{\epsilon \tilde{\tau}} \over {3 \sigma}} \right ).
\end{equation}
The bubble wall action is then
\begin{equation}
	S_E^{Wall} = {\int \sigma \, dA} = 4\pi C^2 \sigma{\int_{-\tilde
	{\tau}_0}^{\tilde {\tau}_0} R^2 (\tilde \tau) \, d{\tilde \tau}}.
\end{equation}
Substituting from Eq.(15) and performing the Euclidean proper time integral
yields
\begin{equation}
	S_E^{Wall} = {{54\pi^2 \sigma^4 C^2} \over {\epsilon^3}}.
\end{equation}
The interior action is
\begin{equation}
	S_E^{Int} = -{\int \epsilon \, dV} = -4\pi C^2 \epsilon
	{\int_{-\tilde t_0}^{\tilde t_0} {\int_{0}^{R(\tilde t)}  R^2 \, dR}
	\, d{\tilde t}} = -{4\pi \over 3} C^2 \epsilon
	{\int_{-\tilde t_0}^{\tilde t_0} R^3(\tilde t) \,d{\tilde t}} .
\end{equation}
The final integration is easily accomplished by converting the
integral over $\tilde t$ to one over $\tilde \tau$ and replacing $R(\tilde t)$
with $R(\tilde \tau)$ from Eq.(15),
\begin{equation}
	S_E^{Int} = -{{81 \pi^2 \sigma^4 C^2} \over {2 \epsilon^3}} .
\end{equation}
Since we are ignoring the gravitational contribution of the false and true
vacuum energy densities, we are free to adjust the zeropoint of energy by
an arbitrary constant, so that the false vacuum energy density, and hence the
false vacuum Euclidean action, will be zero. The vacuum decay parameter $B$ is
then simply the total Euclidean action for the true vacuum bubble nucleated
around a cosmic string or global monopole--the sum of the wall and interior
actions given by Eqs.(17) and (19),
\begin{equation}
	B = {{27 \pi^2 \sigma^4 C^2} \over {2 \epsilon^3}} .
\end{equation}
The bubble action is then seen to be equal to just the usual $O(4)$ action
times the factor $C^2$. Since this factor is less than unity for both the
cosmic string and global monopole metrics, this implies that these topological
defects will act as preferred nucleation sites in the vacuum phase transition.

The deficit angles (and hence deviation of $C^2$ from unity) for global
monopoles and cosmic strings are generally extremely small, unless the
symmetry breaking which creates the topological defect occurs at an energy
scale
very close to the Planck energy. Since the reduction in the Euclidean action
for a bubble nucleated around such a defect is then not too different from
the $O(4)$ value, can nucleation by topological defects ever play an important
role in a vacuum phase transition? The answer, at least potentially, is yes,
as the following demonstrates. The discussion here follows that given for
$O(3)$ nucleation about a gravitational condensation by Samuel and Hiscock
\cite{SaHi2}. The factor $A$ of Eq.(13) is approximated by an appropriate
power of the scale $m$ of the vacuum decay (this value is presumably less than
the symmetry breaking scale $\eta$ which created the global monopole or
cosmic string). For $O(4)$ bubble nucleation, the characteristic time one
must wait for a bubble to nucleate within a box of volume $V$ is approximately
\begin{equation}
	T_4 = {1 \over {\Gamma_4 V}} = V^{-1} m^{-4} \exp {(B_4)},
\end{equation}
while the time to nucleate one bubble around a global monopole in that volume
would be
\begin{equation}
	T_M = {1 \over {n V \Gamma_M}} = n^{-1} V^{-1} m^{-1} \exp {(B_M)},
\end{equation}
where $n$ is the number density of global monopoles. Similarly, the time to
nucleate one bubble centered on a cosmic string would be
\begin{equation}
	T_s = {1 \over {\ell_s V \Gamma_s}} = {\ell_s}^{-1} V^{-1} m^{-2} \exp
{(B_s)},
\end{equation}
where $\ell_s$ is the length of string in a unit three-volume of the spacetime
(which amounts to the ``area density of string'').
The values of $B_M$ and $B_s$ are given by Eq.(20); they may be written as
\begin{equation}
	B_M = B_s = C^2 B_4 = B_4 (1 - 8 \pi \eta^2).
\end{equation}

Whether topological defect nucleation plays an important role in vacuum decay
is determined by the ratios $T_4/T_M$ and $T_4/T_s$. If the value of either
of these ratios is greater than unity, then topological defects will cause
the vacuum phase transition to occur more rapidly than predicted by empty
space $O(4)$ calculations. Evaluating these ratios yields
\begin{equation}
	{T_4 \over T_M} = n m^{-3} \exp{(8 \pi \eta^2 B_4)},
\end{equation}
\begin{equation}
	{T_4 \over T_s} = \ell_s m^{-2} \exp{(8 \pi \eta^2 B_4)}.
\end{equation}
It is now clear under what conditions vacuum decay nucleation by topological
defects might be significant. One would expect the factors $n m^{-3}$
or $\ell_s m^{-2}$ to be much less than unity. For any but Planck scale
monopoles and strings, $8\pi\eta^2$ will also be substantially less than
unity. In order for the ratio of timescales to be greater than unity,
it is then necessary that $B_4$ must be substantially greater than unity. This
will be the case only in severely supercooled phase transitions. Thus, as might
be expected {\it a priori}, the vacuum phase transitions where nucleation might
play an important role are those where the false vacuum enjoys an
exceedingly long metastable life.

The reduction found here in the Euclidean action follows directly from the
reduction in the volume (and surface area) of the instanton solution caused
by the change in the spacetime metric due to the presence of the topological
defect. Similar (but more complicated) volume and surface effects are known
to reduce the action for nucleation about a black hole or other compact
object ({\it e.g.}, a gauge monopole) \cite{Hi3,MeHi}. In those cases, however,
the reduction in action is only significant if the bubble radius is comparable
to the size of the defect, whereas in the present case, the reduction in action
is essentially independent of the size of the defect
core, so long as the core occupies a negligible fraction of the bubble volume.
This is because the spacetime metrics of a black hole or other compact object
such as a gauge monopole are asymptotically flat, globally approaching
Minkowski behavior at infinity. For the gauge string, the spacetime is
everywhere flat outside the string core; however, the nonzero deficit angle
is a global feature of the spacetime, influencing the volume and surface
area of the instanton even at arbitrarily large distances from the string
core. The spacetime of the global monopole is not even asymptotically flat,
and has a solid angle deficit which is independent of distance from the core.

Other, direct interactions of the defect's scalar fields with the field
involved in
the phase transition will most likely be much more important than the
spacetime effects studied here; however, the effects described here are
universal; they will always reduce the Euclidean action, even when there
is no nongravitational interaction between the defect and the
phase transition fields.

This work was supported in part by National Science Foundation
Grants Nos. PHY-9207903 and PHY-9511794.

\end{document}